\documentclass[useAMS,usenatbib,asm]{mn2e}

\usepackage{graphicx,fleqn,times,color}
\usepackage{natbib}

\title[On galaxy spiral arms' nature as revealed by rotation frequencies]{On galaxy spiral arms' nature as revealed by
 rotation frequencies}
\author[Roca-F\`abrega et al.]{Santi Roca-F\`abrega $^{1}$, Octavio Valenzuela $^{2}$, Francesca Figueras $^{1}$, Merc\`e 
Romero-G\'omez $^{1}$, \newauthor  H\'ector Vel\'azquez $^{3}$, Teresa Antoja $^{4}$, B\'arbara Pichardo $^{2}$\\ \\
$^{1}$ Departament d'Astronomia i Meteorologia and IEEC-UB, Institut de Ci\`encies del Cosmos de la Universitat de Barcelona,\\
     Mart\'i i Franqu\`es, 1, E-08028 Barcelona.\\
$^{2}$ Instituto de Astronom\'ia, Universidad Nacional Aut\'onoma de M\'exico, A.P. 70-264,
   04510, M\'exico, D.F.; Ciudad Universitaria, D.F., M\'exico.\\
$^{3}$ Instituto de Astronom\'ia, Universidad Nacional Aut\'onoma de M\'exico, A. P. 877, 22800, Ensenada, M\'exico.\\
$^{4}$ Kapteyn Astronomical Institute, University of Groningen, PO Box 800, 9700 AV, Groningen, The Netherlands. \\
}

\begin{document}

\date{Accepted 2013 April 12.  Received 2013 April 4; in original form 2013 February 27}
\pagerange{\pageref{firstpage}--\pageref{lastpage}} \pubyear{2013}
\maketitle 
%\end{document}

%\begin{document}
\label{firstpage}

\begin{abstract}
High resolution N-body simulations using different codes and initial condition techniques reveal two 
 different behaviours for the rotation frequency of transient spiral arms like structures. Whereas unbarred disks present spiral 
arms  nearly corotating with disk particles, strong barred models (bulged or bulge-less) quickly develop a bar-spiral structure 
dominant  in density, with a pattern speed almost constant in radius. As the bar strength decreases the arm departs from bar rigid 
rotation and behaves similar to the unbarred case. In strong barred models  we detect in the frequency space 
other subdominant and slower modes at large radii, in agreement with  previous studies, however  we also detect them in 
the configuration space. We propose that  the distinctive behaviour of  the dominant spiral modes can be exploited in order to 
constraint the nature of Galactic spiral arms by the astrometric survey GAIA and by 2-D spectroscopic surveys like CALIFA and MANGA 
in external galaxies. 
\end{abstract}

\begin{keywords}
Galaxy: evolution --- Galaxy: kinematics and dynamics --- Galaxy: structure
\end{keywords}

\section{Introduction}
Since the early seventies, it has been suggested that the dynamics driven by bars
and spirals have profound consequences on the kinematic and structural evolution
of galactic disks \citep[e.g.][]{Miller70, Hohl1971, Athanassoula1980, 
Sellwood86,Friedli93}. More recently, stellar radial migration in disk galaxies has been recognized as a 
critical component of disk galaxy evolution.  This process may drastically alter our view
 of the connection between the present-day phase space and chemical distributions of stars and the processes of 
disk formation and evolution.  
 \citet{Sellwood02} set up the dynamical framework of this process 
through the effects of transient spiral structure, which seems to be a crucial
process to understand  the solar neighborhood observations as suggested
originally by \citet{Wielen1996}.
Authors such \citet{Rok2008}, \citet{Schonrich09}, \citet{Roskar11} and
 \citet{Minchev12},
among others,  revived the study of  spiral arms and bars as triggers of stellar
radial migration. Despite these numerous studies, fundamental questions arise
such as: what is the nature of the spirals? or,
how is their pattern speed related to the motion of the stellar component?
Several models have been proposed up to now, from the classical Tight-Winding Approximation
\citep[TWA, e.g.][]{Binney08} to the mechanisms proposed to account
for self-excited spiral patterns \citep{Toomre90,Bertin96,Sellwood00,Donghia12}, or the
 manifold theory \citep{Romero07,Tsoutsis09,Athanassoula2012}.\\
In the quest of dynamical models not limited to quasilinear 
 approximations or steady state, N-body simulations have been used to understand the origin and
evolution of spiral structures.  After  pioneering studies like \citet{Miller1979,Sparke1988,Rautiainen99}, only recently appeared 
another boom of papers likely as a result of  progress in the computational resources and codes. This re-ignition of  simulations 
based studies opened-up an interesting debate about the possible corotating nature of spiral patterns  with disk particles
 \citep[e.g.][]{Quillen11,Grand2012a,Grand2012b,Minchev12,Donghia12,Baba12}. As an example,
\citet{Grand2012a,Grand2012b} computed collisionless N-body and smoothed
particle hydrodynamics (SPH) simulations of disk galaxies and illustrated that transient spiral features appear to corotate with the 
disk.
 \citet{Comparetta12} also suggested that some short lived features arising from constructive interference between longer lived modes,
 i.e. fast bar and slowly moving spiral pattern modes, can be nearly corotating with the disk. 
This corotating nature would have consequences on the stellar radial migration mechanisms \citep{Grand2012a}.\\
  From the observational point of view, the nature of spiral arms is also far from being clear \citep{Sheth10,Foyle11,Ferreras12}. 
Nowadays, only weak observational constraints are available to answer these questions. Some constraints come out from  direct 
measurement of the
  rotation frequencies radial variation in external galaxies \citep[e.g.][]{Meidt09}, a method with a strong potential but currently 
applied only to a handful of galaxies, or from indirect measurements as the one proposed by \citet{Martinez09}. Other constraints to spiral arms' nature are based on small scale stellar  kinematic substructure
 analysis in the Milky Way (MW) disk \citep[e.g.][]{Antoja11,Antoja12}.
Furthermore, although it is now commonly accepted that the MW is a barred galaxy, it is not clear  how the arms and  bar are related or
 if they are connected at all. The rotation frequency of both, bar and spiral arms, seem to show different  
 values, none in mutual  corotation or in corotation with galactic disk material, leaving unclear spiral arms' nature 
\citep{Martos04,Gerhard11}. It is fair to say that currently,  galactic arms'  pattern speed estimations are mostly model dependent
 \citep{Martos04}. The situation is also not clear for
 external galaxies \citep{Buta05}.  The interpretation of observational results is based on previous theoretical studies and also
 on numerical simulations \citep{Sparke1988,Rautiainen99}. Some of these studies adopted
 simplifications like 2D N-body models and a rigid halo or no halo at all. It is not clear if such assumptions may affect the 
generality of their conclusions as is suggested by \citet{Athanassoula02} for the case of bar growth.  It is also not  obvious 
if the  corotating nature of spiral arms recently suggested by \citet{Grand2012a} is valid also in models with different structure 
(i.e. models with strong/week bar and/or bulge).  Furthermore, classical methods to estimate disk modes' rotation frequencies  like 
time Fourier spectrograms applied to simulations, have been claimed to suffer from biases in models with multiple or week spiral arms,
 hampering estimations of arms' pattern speeds \citep{Grand2012a}.\\  
In this paper we analyze 3D galaxy models, with different stellar/dark structure, using live halos and with enough mass,
 force and time resolutions to accurately describe the internal disk kinematics. We performed some testing on the results
 dependence on codes and initial conditions techniques. It should be emphasized that the N-body simulations presented here
 are purely stellar. From the first attempts to simulate gas in barred
 galaxies \citep{Sanders76}, it has become clear that gas influences disk stellar dynamics, by even changing the live time of bars
 \citep{Bird12,Dimatteo13}. Gas makes disks to show more complicated and well-defined
 morphologies. As an example, inner rings, such the Galactic Molecular Ring observed in our MW \citep{Clemens88},
 have been the subject of numerous observational \citep{Buta04} and theoretical investigations \citep[e.g.][]{Byrd06,Romerogomez11}.
 The influence of the gaseous component and its subgrid physics in our simulations is now under investigation using the ART code in
 the hydrodynamic version \citep{Kravtsov2003,Colin11}.\\
The aim of the first study presented here is to revisit the possible correlation between spiral arm kinematics
 and their nature using the purely stellar component. In Section ~\ref{sec:nbody} we describe our 
N-body simulations, carried out using
two well known  N-body codes, ART and GADGET3. In Section ~\ref{sec:methods}, we describe the techniques used to
 derive the rotation frequencies, whereas the results and conclusions are presented in Section ~\ref{sec:results}
 and ~\ref{sec:disc}, respectively.

\begin{table}
\centering
\begin{tabular}{|c||c|c||c||c||c|}
\hline 
\hline 
Parameter &  B1/5  &  G1 & U1/5 \\ 
\hline 
\hline
Disk mass ($10^{10}$ $M_{\odot}$) & 5.0 & 6.01 & 3.75\\
\hline
Halo mass ($10^{12}$ $M_{\odot}$) & 1.38 & 0.66 & 1.5 \\
\hline
Disk exp. length $R_d$ (kpc) & 3.86 & 3.0 & 4.0 \\
\hline
Disk exp. height $Z_d$ (kpc) & 0.2 & 0.2 & 0.2 \\
\hline
Halo NFW $R_d$ (kpc) & 29.19 & 14.4 & 16.61 \\
\hline
Halo concentration & 10 & 10.4 & 18 \\
\hline
Halo DM species & 6/7 & 1 & 6/7 \\
\hline
N$_{*disk}$(+N$_{*bulge}$) ($10^6$) &  1.0/5.0 &  0.5(+0.5) & 1.0/5.0 \\
\hline
$N_{eff}$ ($10^7$) & 2.86/13.8 & 0.15 & 4.1/20.0 \\
\hline
Min. time step ($10^4$ yr) & 3.2/1.6 & 4.4 & 7.9/3.1 \\
\hline
Spatial Resolution (pc) & 44.0/11.0 & 35.0 & 11.0/11.0 \\
\hline
Total integration t (Gyr) & 4.6/2.8 & 1.4 & 3.2/2.8 \\
\hline
\hline
\end{tabular}
\caption{Parameters of the simulations: Column indicated by B1/5 presents the parameters for models with 1 million (B1) 
and 5 million (B5) star particles in the disk, simulated using ART code,
 idem for unbarred models U1/5 in the third column and for the barred model G1, using GADGET3 code, in the second column.}
\label{tab:1}
\end{table}

\begin{figure*}
\centering
\includegraphics[scale=0.21]{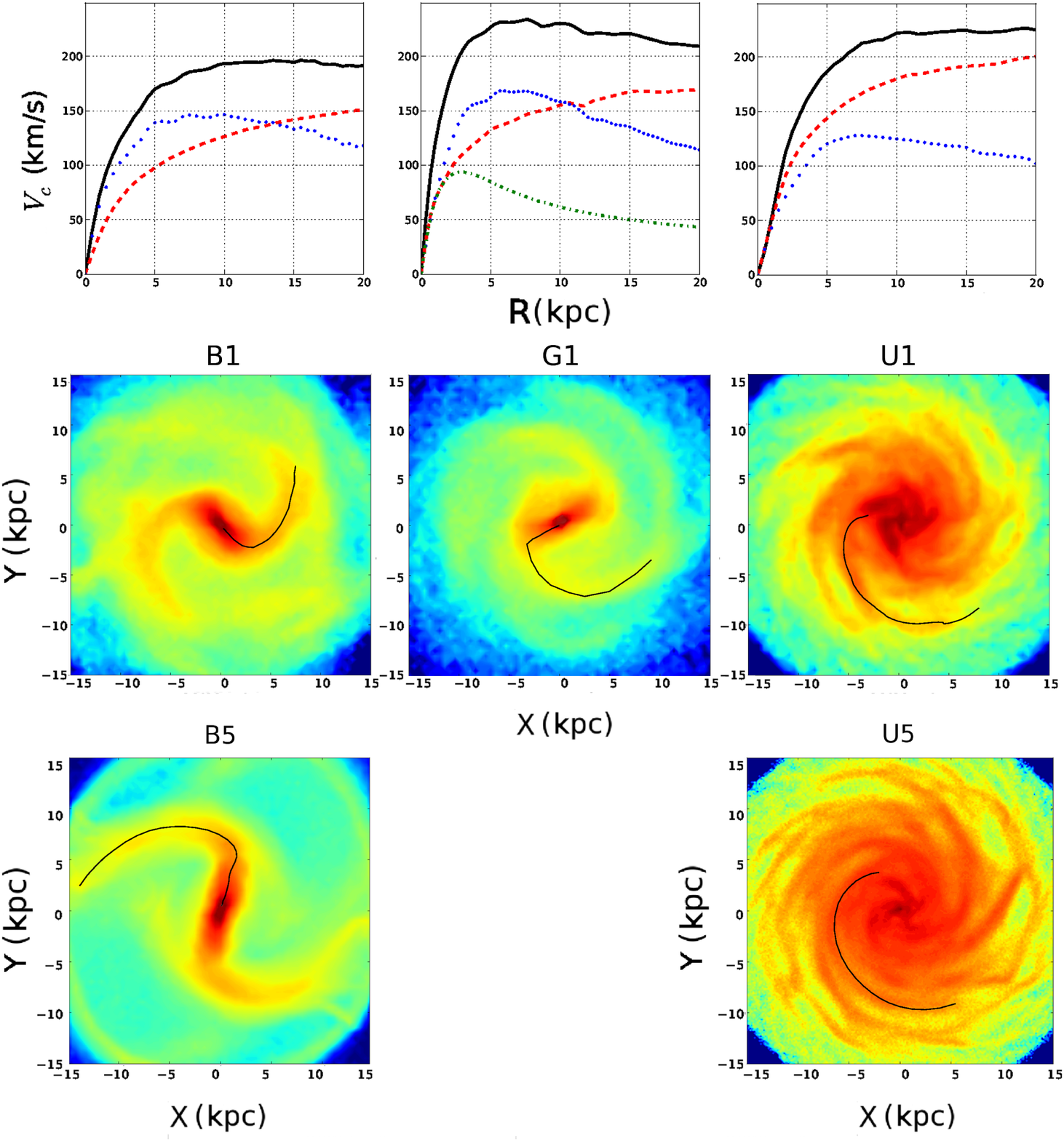}
\caption{Structure of the models. Top: Initial circular velocity of models B (left), G (center) and U (right)
, computed from the potential field gradients using TIPSY package.
 We show the total rotation curve (black solid), disk (blue dotted), halo 
(red dashed) and bulge (green dot-dashed) contributions. Center: Density distribution of the models B1 (left),
 G1 (center) and U1 (right) after $900$ Myr of evolution. Bottom: Density distribution of the models B5 (left) and
 U5 (right) after $900$ Myr of evolution. The black solid line shows the locus of the spiral arms derived using Fourier
 analysis ($m=2$ for B and G models, $m=4$ for U). Spiral structure rotates clockwise in all models.}
\label{fig:fig1}
\end{figure*}

\section{Models and simulations}\label{sec:nbody}

\begin{figure*}
\centering
\includegraphics[scale=0.15]{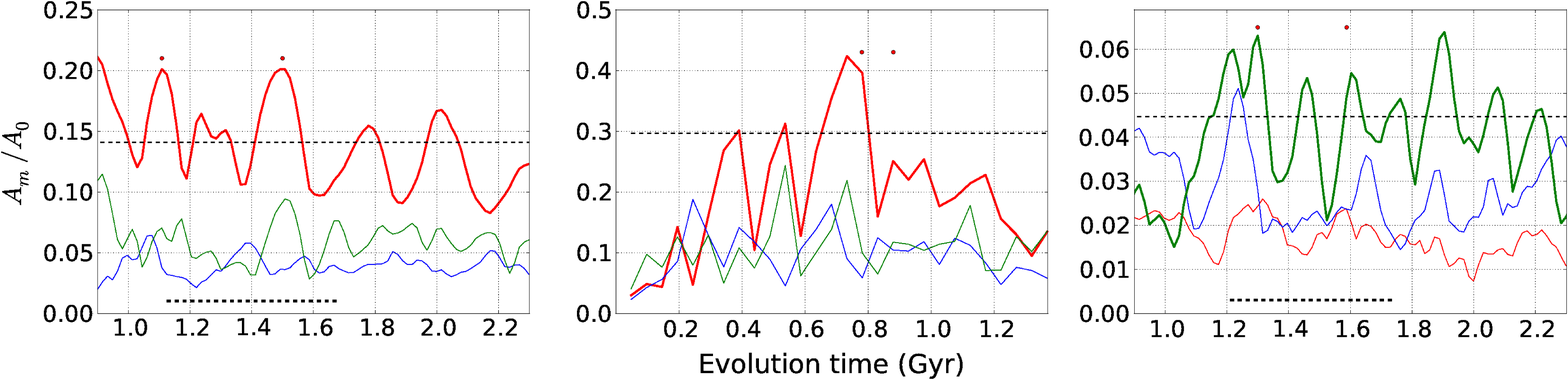}
\caption{Amplitude evolution of disk modes. Fourier amplitude for modes 2 (red), 3 (blue) and 4 (green) as a function
 of time, averaged for radius between $4$ and $10 {\rm kpc}$ of the barred (B1, left), the bulge-barred (G1, center) and the 
unbarred (U1, right) models 
(notice change in vertical scale). Thin black dashed lines indicate a threshold in amplitude used to compute Figure~\ref{fig:fig3}
using SFpFD method. The black dots indicate the snapshots for which the kinematic analysis is shown in Figure~\ref{fig:fig4} using
 SFpFD method, and thick black
dashed lines indicate the temporal range used to compute spectrograms in Figure~\ref{fig:fig5}.}
\label{fig:fig2}
\end{figure*}

We have performed collisionless N-body simulations with ART \citep{Kravtsov1997}  and GADGET3 
\citep[last described in ][]{Springel05} codes.
Tests of consistency of both codes, applied to dynamics of barred galaxies can be found in
\citet{Valenzuela2003}, \citet{Colinetal06} and in \citet{Klypin2009}.
We present here three sets of fully self-consistent models, all of them  with a live exponential 
disk and live dark matter halo with the NFW \citep{Navarro97} density profile (see Table ~\ref{tab:1}). The live halo ensures disk-halo angular 
momentum exchange, which plays
 an important role in the formation and evolution of bars as discussed by \citet{Athanassoula02}.
We  simulated barred (B) and unbarred (U) models with the aim to compare spiral arms potentially triggered by a bar against arms triggered by other mechanisms.
  Models B and U were simulated using the ART code. To generate initial conditions of these models we 
 used the Jeans equation moments method as introduced by \citet{Hernquist93}. We have also used  a multimass method to sample the 
halo particle distribution, which allows us to obtain similar results as using a higher number ($N_{eff}$) of particles, minimizing 
two body scattering as discussed in \citet{Valenzuela2003}. 
Barred models have a stellar disk and total mass similar to the one observed for the MW with additional initial parameters as proposed by 
\citet{Colinetal06} (model K$_{cb}$). As discussed in \citep{Klypin02}, the final properties of this model, after rescaling, can reproduce the observed quantities for
the MW. This rescaling process has not been applied since it does not affect the disk kinematic properties analysed in this paper.
Unbarred models have a smaller disk and a massive and 
highly concentrated halo. Additionally we have used the GADGET3 code to simulate barred models, labeled G, including a bulge component 
with a
 Sersic profile ($R_b=1.75$ kpc, $M_b=8.57\cdot10^{9}$ M$_{\odot}$). For the bulged models' initial
 conditions  generation we have used the code described in  \citet{Widrow05}. The number of disk particles in our models 
 range from one to five million particles (see Table ~\ref{tab:1}).\\
As expected \citep[e.g. ][]{Ostriker73}, the models with a relatively dominant disk (B and G) rapidly generate a bar with trailing 
spiral 
arms while unbarred models (U) do not , at least  in the first $\sim3 {\rm Gyrs}$. In the first case, we have a 
similar halo and disk contribution to the circular velocity inside a radial exponential length and, as described in 
\citet{Valenzuela2003}, the self-gravity of disk density fluctuations (disk modes) dominates making the instability grow 
and generate a bar (Figure~\ref{fig:fig1}, left and middle panels). This bar mode induces the formation of a bisymmetric spiral 
structure apparently connected to the bar ends \citep[e.g.][and references therein]{Binney08}. There is still controversy on the 
nature of these
bisymmetric spirals (resonant coupling, manifolds, or others), but what is evident in all our models  is that these structure
 dominates in density.
For the unbarred model (Figure~\ref{fig:fig1}, right) the halo contribution to the potential is higher 
and the disk modes cannot grow so easily.  Although in both cases the
initial velocity dispersion is low ($Q=1.2$), the higher halo mass concentration 
in the unbarred model prevents disk from having a dominant bisymmetric mode in the first Gyrs of evolution, i.e. 
other modes grow forming a trailing $3$-$4$ armed structure. Some other
structures with lower density appear in the external regions of these simulations (both barred and unbarred). In the next section we
discuss their imprints on the frequency space. As expected for MW like galaxies, we also note that rotation curves in our set of simulations are 
rather flat (see top panels of Figure~\ref{fig:fig1}).\\
In all our models spiral arm structures are observed for at least $2$-$3$ ${\rm Gyrs}$. We show in Figure~\ref{fig:fig2} the temporal
 evolution of the spatial Fourier modes for the three models with one million particles in the disk.
 We see that dominant spiral modes have a recurrent nature with periodicities of less than one galactic rotation. The amplitude of the
spiral arms in the unbarred case  is significantly lower than in the barred models.\\
In Figure~\ref{fig:fig1} we observe that some  parameters of the bar and
 spirals are different from B1 (one million disk particles) to B5 (five million) models. These differences are arising from the fact that
the number of particles, spatial and time resolution in B5 model are much better than in B1, i.e. we are resolving smaller wavelength 
disk modes in the B5 case than in B1. These modes interact with the disk and then, due to the high non-linearity of the system, lead 
the evolution to a slightly different configuration (e.g. bar length and speed). However global quantities like circular velocity and
 density profiles are more robust to such effects. The situation is well known and it has been reported in previous works \citep{Sellwood09,Klypin2009}.
  A convergence study of this and other models will be presented in Roca-Fabrega et al. (in preparation), however, as will be seen
in next sections, the main results of 
our simulations are robust across changes in numerical parameters:  a bar-arm structure which is dominant in density plus external
 and weaker arms are formed in barred disks, while in barless models low amplitude arms are found. 
In the next section we will analyse the kinematics of such spirals structures.\\   

\section{Overdensities and rotation frequencies}\label{sec:methods}

We use spatial Fourier analysis azimuthally averaged in order to trace the density peak of the spirals and the bar
  \citep{Valenzuela2003}, working in cylindrical shells equally spaced in galactocentric distance.
 Figure~\ref{fig:fig1} shows an example of how well the spatial Fourier method
 traces the peak overdensities up to the end of the dominant structures (m=2 for barred cases and m=3,4 for unbarred ones).
 We also used a density peak method as the one used in 
\citet{Grand2012a,Grand2012b} to test that the results are method independent. 
 Once the locus of the spiral is derived, we use both, finite differentiation among three consecutive snapshots and the classical 
spectrograms
 method \citep{Sellwood86} to compute the rotation frequency.\\
 The advantages of using Spatial Fourier plus Finite Differentiation (hereafter SFpFD) are that it allows us to compute the rotation 
frequency of a
 single known mode $m$ structure and for a single time instant. As a consequence, SFpFD is able to show us how the
 structures evolve with time. On the contrary the spectrogram method \citep{Sellwood86} needs
 to be applied to a large time interval due to the Nyquist frequency limitation.
 Because of that, the results from spectrograms method may be contaminated by recurrent arms sequences with low or negligible 
spiral amplitude due to the
 transient nature of the structures. This drawback is already discussed by \citet{Grand2012a}.\\
A weakness of using SFpFD, independently of having two or more spiral arms, appears when two coexisting structures of the
same
 Fourier mode at the same radius are present. In this case, the method results in an unique structure placed at the average angle.
 Thus, we have to control these cases to avoid a bias in the derived rotation frequency. In contrast 
spectrograms can find discrete rotation frequencies from structures coexisting at the same radius without any problem.
After carefully weighting the pros and cons we state that both methods (SFpFD and spectrograms) are complementary.\\

\begin{figure*}
\centering
\includegraphics[scale=0.23]{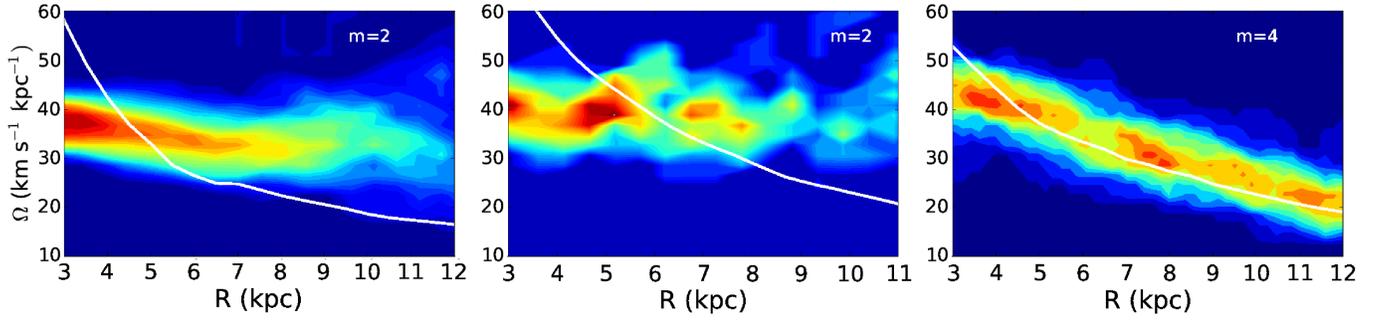}
\caption{Rotation frequencies as a function of radius calculated using SFpFD method for dominant mode across time in models B1 (left,
 G1
 (center) and U1 (right) (see Figure~\ref{fig:fig2}). Here we plot the frequency density map of the rotation frequencies computed
 using all time instants when the spiral
 arms' amplitude is above 70\% of the maximum mode amplitude (dashed line in Figure~\ref{fig:fig2}). 
 Circular frequencies of disk particles are indicated as solid white lines and have been computed in an intermediate instant of the 
analyzed time interval. The length of the bar is $\sim4.5 {\rm kpc}$ and $\sim5 {\rm kpc}$ for B1 and G1 models respectively.}
\label{fig:fig3}
\end{figure*}

\begin{figure*}
\centering
\includegraphics[scale=0.16]{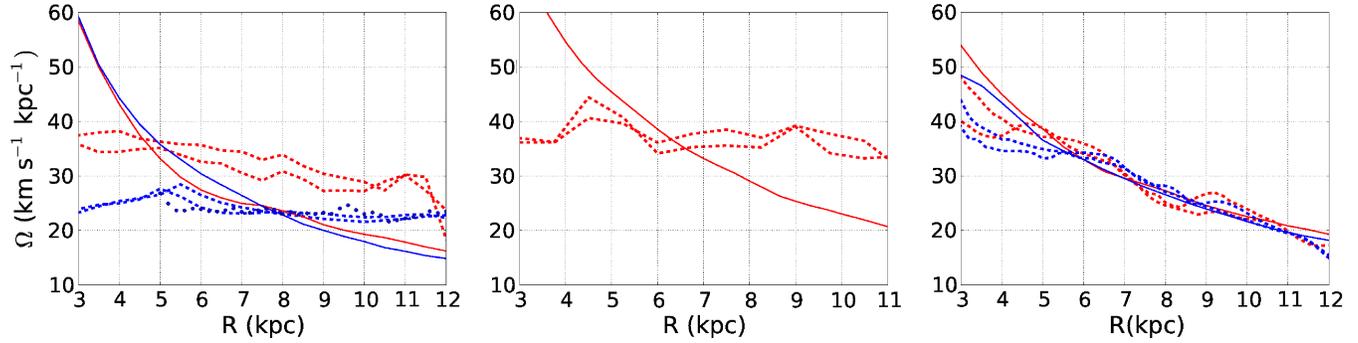}
\caption{Rotation frequencies as a function of radius calculated
 using SFpFD method for selected time instants when the amplitude of spiral arms is maximum (black dots
in Figure~\ref{fig:fig2}). 
 All the rotation frequencies analysis of barred (B1, and B5 left), the bulge-barred (G1, center) and the unbarred (U1, right) has been 
computed taking cylindrical shells of $0.5 {\rm kpc}$ width. In red dashed we show results obtained
for models with 1 million disk particles. In blue dashed,
 those of the ART models, with 5 million disk particles (left and right panels respectively, for models B5 and U5) when the amplitude 
of the spirals 
arms is maximum (0.75 and 1.12 Gyr). Circular frequencies of disk particles are indicated as solid lines (red for models B1 and U1 and 
blue models B5 and U5).
Additionally, blue dots in the left panel show the results of applying a density peak method similar to one used in \citet{Grand2012a} to B5 simulation, 
to find the spiral structure (a detailed description
of this method will be included in Roca-F\'abrega et al. in preparation). The dispersion on rotation frequency profiles due to its
computation at several time steps when the amplitude of the dominant mode and its behaviour can slightly change, is shown at Figure~\ref{fig:fig3}.
 The length of the bar is $\sim4.5 {\rm kpc}$, $\sim7.0 {\rm kpc}$ and $\sim5.0 {\rm kpc}$ for B1, B4 and G1 models 
respectively. The differences in bar properties between models B1 and B5 are discussed in Section~\ref{sec:nbody}.}
\label{fig:fig4}
\end{figure*}

\begin{figure*}
\centering
\includegraphics[scale=0.29]{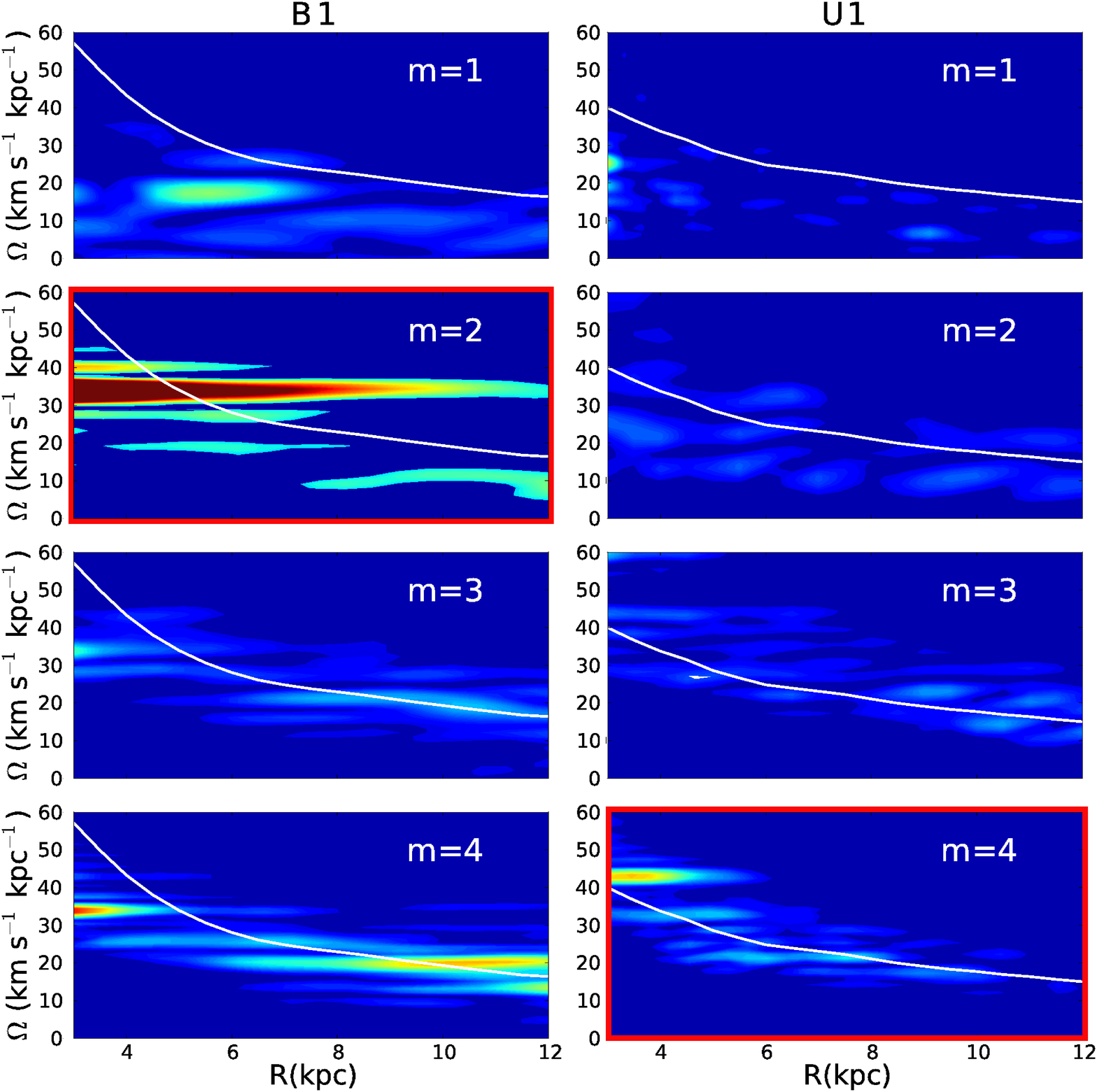}
\caption{Spectrograms for the barred model B1 (left) and unbarred U1 (right) obtained using \citet{Sellwood86} method.
 We show the Fourier component for the m=1,2,3 and 4 modes in a window spanning 0.5 Gyr centered at 1.4 Gyr (left) and 1.45 Gyr (right),
 (see dashed dark lines in top panels of Figure~\ref{fig:fig2}), and with a Nyquist frequency $\sim100-150 {\rm km s}^{-1} {\rm kpc}^{-1}$. The
 y axis is the angular frequency in ${\rm km s}^{-1} {\rm kpc}^{-1}$ and the
 x axis is radius in kpc. Overplotted in white we show the rotation curve of disk particles computed for an intermediate instant of 
the analyzed time interval.}
\label{fig:fig5}
\end{figure*}

\section{Spiral arm rotation frequencies}\label{sec:results}

We perform a first kinematical analysis on the dominant modes in density shown in Figure~\ref{fig:fig2}, $m=2$ for the barred (bisymmetric spiral)
 and $m=4$ for the unbarred models (4 armed structure).
As can be seen in Figure~\ref{fig:fig1}, the dominant density structures extend up to $10-11$ kpc. 
In Figure~\ref{fig:fig3} we present the rotation frequency
 curves  for the three models with one million disk particles computed using the SFpFD method. It includes all timesteps where we can 
ensure the spiral arm is well formed. We empirically
 establish that a spiral is well formed when the amplitude of the dominant mode is above 0.7 times the maximum 
value of this mode in the range we study (see thin black dashed lines in Figure~\ref{fig:fig2}). These density maps are constructed from the
 superposition of all the
 rotation frequency curves at timesteps when the amplitude of the mode is above the mentioned threshold. These figures show 
significant differences between 
barred and unbarred morphologies. Note how in the barred models (B1 and G1, computed using ART and GADGET3, respectively) the spiral 
pattern rotates almost as a rigid body, while in 
unbarred models the rotation frequency curves lie on top of the rotation of the disk particles, 
resulting in a spiral mode corotating with the disk.\\
A more exhaustive analysis is shown in Figure~\ref{fig:fig4}, considering models with a higher number of disk particles and a different technique 
for the detection of the spiral structure.
Instead of working with density maps, here we plot single timestep
 curves computed using SFpFD for models with both one and five million disk particles. For the B and G models we avoid
 the central part of the bar, for U the central complex region where the 3-4 armed structure converges and in both we also 
avoid the external regions where the number of particles is too low. The red-dashed lines correspond to two instants of
 the one
 million disk particle models (included in the density plot of Figure~\ref{fig:fig3}) shown as red dots in Figure~\ref{fig:fig2}.
More important, the blue-dashed lines correspond to models with five million disk particles, both barred and unbarred, integrated 
using ART. Furthermore, the blue-dotted 
line corresponds to the barred ART five million model for which the spirals have been detected using the density peak method (similar 
to the one
 in \citet{Grand2012a,Grand2012b}. Note how the conclusions reached in Figure~\ref{fig:fig3} are well corroborated. We can see spiral 
patterns rotating almost as a rigid body in all
our barred models, both using ART and GADGET3, using different techniques for the spatial detection and considering a different number
 of particles. We have also verified that simulations 
with $2\cdot10^5$ disk particles show the same behaviour (not included here). A slightly decreasing rotation frequency with radius is 
observed only for the 
ART model with one million disk particles (B1). Even though, this behaviour is completely different from all our ART unbarred models, 
both using one million and five million disk particles.\\
To ensure that our results are not dependent of the method used to compute the rotation frequencies, in Figure~\ref{fig:fig5}, we 
perform one last test using, in this case, the classical 
spectrogram method in B1 and U1 models with a Nyquist frequency of $\sim 100-150$ km s$^{-1}$ kpc$^{-1}$. The
 panels framed in red correspond to the dominant modes discussed above, and they are to compare with the left and right panels of 
Figure~\ref{fig:fig3}.
 Note that, again, the barred model presents a dominant mode that rotates as a rigid body, while in the unbarred case, the structures 
corotate with the disk particles. Two features observed in this dominant modes deserve special attention. First, in the
 barred model, a well defined structure is observed at a radial range of $11-14$ kpc (only the beginning of the structure is shown in 
the figure). It rotates slower than the disk 
with a frequency of about 
 $8$ km s$^{-1}$ kpc$^{-1}$. As mentioned in Section 3, here we note one of the advantages of the spectrograms. They are able to
 detect multiple spiral patterns at a given radius. 
Using SFpFD, we have checked that the low frequency structure corresponds to one of the lower density structures observed in the 
outer radii of the center-left panel of Figure~\ref{fig:fig1}. 
The complex link with the dominant structure will be discussed in a forthcoming paper. As it is known, these multiple pattern speeds
 are common in spiral barred galaxies both in simulations and observations \citep[e.g.][]{Masset97,Buta11}. Secondly, and less
 important, we can appreciate a flat structure in the rotation frequency in the central region of the unbarred 
model (R $<$ $4$ kpc, bottom right panel of Figure~\ref{fig:fig5}), which could remind the signature
 of a weak bar mode. This is a misinterpretation since it corresponds to a $m=4$ mode, and, actually, the amplitude of the $m=2$ mode
 in this region is less than 2\% (see Figure~\ref{fig:fig2}).\\
Figure~\ref{fig:fig5} also includes the spectrograms for the subdominant modes. As in Figure~6 of \citet{Grand2012a} we find all these
 subdominant
 modes clearly corotating with the disk. This behaviour is also observed in other studies as in \citet{Minchev12} (Figure~8), although
 not 
discussed there, while they are centered on the study of resonant coupling.\\

%\begin{figure*}
%\centering
%\includegraphics[scale=0.3]{fig6.eps}
%\caption{In red, mean and dispersion of the rotation frequency computed at all time intervals when the bar
% and the spirals are well formed ($0.6$ Gyr to $1.8$ Gyr of evolution) using SFpFD method, for B5 simulation. In green we show the
% mean disk rotation frequency in the analysed time interval.}
%\label{fig:fig6}
%\end{figure*}
%\begin{figure*}
%\centering
%\includegraphics[scale=0.5]{fig6_1.eps}
%\caption{Rotation frequency density map, same Figure as Figure \ref{fig:fig3} but now for B5 simulation, computed at all time
% intervals when the bar and the spirals are well formed ($0.6$ Gyr to $1.8$ Gyr of evolution) using SFpFD method. \textcolor{red}{Debemos
%escoger si mostramos esta, la anterior o ninguna de las dos.}}
%\label{fig:fig6_1}
%\end{figure*}

\section{Discussion and Conclusions}\label{sec:disc}

The rotation frequencies of the spiral modes in barred and unbarred models,
 integrated with different N-body codes (ART and GADGET), and analysed with different  techniques
 (spectrograms and finite differences), present a well defined different behaviour.  Whereas unbarred 
models show transient arms corotating with disk particles, in good agreement with those recently reported by \citet{Grand2012a} 
and \citet{Baba12}, 
 barred models  deserve further comments.  As shown in Figures~\ref{fig:fig3} and \ref{fig:fig4},
 barred models present a spiral pattern speed almost constant in radius for all the range where the spiral structure is dominant
 in density. These results are consistent with those of models I, II and III in \citet{Rautiainen99}, computed using
 a simplified 2D model and with rigid or no halo. We want to emphasize that whereas in those early models it is difficult to
 detect spiral structures in the configuration space, particularly in the external regions, in our simulations with 1 and 5 million
 disk particles, the spiral structure is clearly identified (see Figure~\ref{fig:fig1}), allowing us to
 use the arm phase in order to test claimed biases in the derivation of the rotation frequencies using spectrograms (see Section 
\ref{sec:methods}).\\
It is important to compare our results with those recently obtained by \citet{Grand2012b}. The authors  have analyzed
 N-body/SPH simulations of isolated barred galaxies concluding  that,  spiral arms'  pattern speed 
 decreases with radius, closely in corotation with disk particles.
 \citet{Grand2012b} computed the rotation frequencies averaging the values obtained for several snapshots over two time
 intervals during spiral arm evolution, one when the bar is well defined in their images ($\sim1$ Gyr of evolution, see their Figure~9)
 and the other when the bar has significantly weakened ($\sim1.5$ Gyr, Figure~10). In both cases, the spirals pattern speed is almost 
corotating
 with  disk particles. The previous result is at odds to what is obtained in our study. However they also noticed  a small offset  of
  the spiral arms' rotation frequency 
profile from being corotating with disk particles, particularly when the bar is stronger, suggesting that further analysis is required.
 Here
 we have analyzed the correlation between the strength of the bar and the spiral rotation frequency in our
 B1 ART model, a Milky Way like simulation. With such a purpose, we have computed  arms' rotation frequency during the first 
evolutionary stages,  
 when the bar is still growing and weak ($A_2/A_0 < 0.1$) at about $0.3$ Gyr of evolution.
 Only when the bar is weak we observe that the external m=2 mode (bisymmetric arm) rotation frequency is close to disk corotation.
 After few galactic
 rotations, when the bar has fully formed ($A_2/A_0 > 0.4$)  (as shown in Figure~\ref{fig:fig2} left panel),
 the rotation frequency becomes almost constant,  approaching the bar rigid body rotation, if any with a small decrement in radius
 (see Figure~\ref{fig:fig3} and \ref{fig:fig4}, left and central panels). This result is qualitative in agreement with the statement 
discussed by \citet{Grand2012b}, unfortunately we do not have information of bar amplitude in \citet{Grand2012b} simulations, and 
their models assume a gas component making harder a further comparison.  We conclude that in barred models 
the spiral rotation frequency  approaches the bar rigid body rotation with the increment of bar strength.  \\
Furthermore, our simulations, with a high number of disk particles, allow us to analyze the behaviour at larger radii $\sim12-14$ kpc. 
As can be seen in Figure~\ref{fig:fig5} (m=2 mode, left panel second row), the spectrograms method applied to the B1 bar model shows 
an external structure,
 at radii larger than R= 11 kpc,  that rotates slower than the density dominant mode. This external structure has a lower amplitude, 
lower 
rotation frequency (less than $10$ km s$^{-1}$ kpc$^{-1}$), it is tightly  wound, and
 it is recurrently connected/disconnected with the inner and faster spiral (see movie\footnote{This movie can be downloaded 
from http://www.am.ub.edu/$\sim$ sroca/Nbody/movies/B1.mpeg and it shows the density evolution of model B1, spanning from 0.1 to 3.1 
Gyrs. High density regions ($\sim3.7$ M$\odot/$pc$^3$) are shown in dark-blue colors while zero density ones are in white.}). Other authors 
\citep[i.e][]{Sparke1988}
 have reported similar structures but apparently they detected them only in the frequency space, probably due to the statistical 
fluctuations given
by their number of particles (see however figure~16 in \citet{Rautiainen99}). The lack of detection in configuration space led the
 authors to conclude that inner
fast rotating structures correspond only to the bar structures while the slower and external ones to the entire and unique spiral 
mode.
 Here we confirm the detection and the rotation frequency in configuration space (see Figure~\ref{fig:fig1}) for all these structures,
 and that the inner fast 
rotating structure includes both:  a bar and an inner spiral structure apparently connected. The slow rotating mode is  an external 
spiral mode.  We stress that we also find both structures using the SFpFD method in configuration space.
 In conclusion, at 
large radii, and for m=2, we do see the external slow and low amplitude arm coexisting with the dominant spiral mode which is most of
 the
 time connected to the bar.\\
The results presented here for the barred case are also compatible with the observations recently reported by \citet{Meidt09}. 
They derive rotation frequencies in four spiral galaxies: two strong barred, one unbarred  and one with rings. They find that in the
 two strong barred cases the inner bisymmetric spirals rotate with the same  frequency as the bar. Furthermore,  one of
 the barred cases present an external armed structure rotating with a much lower frequency (see their Figures~8 and 12) as it is 
observed in our B1 barred model. For the
 other cases they find multiple pattern speeds that can be or not in a resonance coupling situation. It is important to mention that
 at least one of their barred galaxies is likely in interaction (M101). As discussed by the authors, the limiting observational accuracy - precision in the radial binning and frequency sampling - do
 not allow them to 
confirm or refute the small radial decrement in rotation frequency shown in our Figure~\ref{fig:fig3}, therefore they conclude 
 that arms in their study are rigid body rotators.  Further observations will be required to confirm  this tendency.\\
From the simulations we have performed  it is clear that a difference in the spiral arm kinematics exists if they are triggered by 
a bar
 or by another mechanism, and also  that barred models seem to show at least two kinds of spiral arms. Further analysis of the arm
 nature, and
 numerical convergence tests  will be discussed in a forthcoming paper.\\
 
 As a summary we confirm  that:
\begin{itemize}
\item The dominant spiral mode (m = 2) in strong barred models is most of the time connected to the bar. Its rotation frequency
 is near to the
bar solid body rotation, with a small decrement with radius which is more important as the bar amplitude decreases.
 Arms are corotating with disk particles only for very weak bars.
\item Although the dominant spiral disk mode in  barred simulations is the one connected with the bar, we observe at least one 
subdominant slow and winded mode at large radii, using either SFpFD, spectrograms (see Figure~\ref{fig:fig5}) and density maps.
\item In unbarred models the spiral structures are corotating with  disk material, in agreement with previous results 
 \citet{Grand2012a}.
\item Our results are robust to changes in numerical parameters (time step, spatial resolution, number of particles), mode analysis techniques 
 (SFpFD, spectrograms, density peak) and also to changes in  numerical codes (ART and GADGET3).
 \end{itemize}
 
A natural question is to ask which is the situation for the MW: Do the traditional spiral arms correspond to the mode
coupled with the bar? or instead correspond to the modes we observed at larger radii? or both? Planned and current surveys
 measuring stellar kinematics and distances inside our Galaxy as Gaia (ESA) or APOGEE (SDSS) will open up the possibility of 
direct estimation through methods like the one proposed by \citet{Tremaine84}.   Spectroscopic high resolution surveys of external
 galaxies like CALIFA \citep{Sanchez2012} or MANGA(AS3/SDSSIV)  will contribute to test our predictions using stellar kinematics.\\

%______________________________________________________________
%\section*{Acknowledgments}
%\textbf{We thank the referee for a constructive report. }
We thank A. Klypin, A. Kravtsov and V. Springel for providing us the numerical codes 
and L.~M. Widrow for providing the code to generate the initial conditions. We thank HPCC project and T. Quinn for the implementation of
TIPSY package. Finally we also thank Luis Aguilar, Ivanio Puerari and Gene G. Byrd for his helpful comments on this work. 
This work was supported by the MINECO (Spanish Ministry of Economy) - FEDER through grant AYA2009-14648-C02-01, AYA2010-12176-E, 
AYA2012-39551-C02-01 and CONSOLIDER CSD2007-00050. SR was supported
 by the MECD PhD grant 2009FPU AP-2009-1636.
Simulations were carried out using Pakal, Abassi2 and Atocatl at IA-UNAM, and Pirineus at CESCA.

%______________________________________________________________
% BIBLIOGRAFIA

\bibliographystyle{mn2e}

% definiciones solo para el MNRAS para ApJ no ponerlas
\def\apj{ApJ}
\def\apjl{ApJ}
\def\aj{AJ}
\def\mnras{MNRAS}
\def\aa{A\&A}
\def\nat{nat}
\def\araa{ARA\&A}
\def\aap{A\&A}

%\bibitem[Krumholz 
%\& Thompson(2012)]{Krumholz2012} Krumholz, M.~R., \& Thompson, T.~A.\ 2012, arXiv:1203.2926 

\bibliography{biblio}

\begin{thebibliography}{}

\bibitem[\protect\citeauthoryear{{Antoja}, {Figueras}, {Romero-G\'omez},
  {Pichardo}, {Valenzuela} \& {Moreno}}{{Antoja} et~al.}{2011}]{Antoja11}
{Antoja} T.,  {Figueras} F.,  {Romero-G\'omez} M.,  {Pichardo} B.,
  {Valenzuela} O.,    {Moreno} E.,  2011, \mnras, 418, 1423

\bibitem[\protect\citeauthoryear{{Antoja}, {Helmi}, {Bienayme} \& {et
  al.}}{{Antoja} et~al.}{2012}]{Antoja12}
{Antoja} T.,  {Helmi} H.,  {Bienayme} O.,    {et al.} 2012, \mnras, 426, L1

\bibitem[\protect\citeauthoryear{{Athanassoula}}{{Athanassoula}}{1980}]{Athana%
ssoula1980}
{Athanassoula} E.,  1980, \aa, 88, 184

\bibitem[\protect\citeauthoryear{{Athanassoula}}{{Athanassoula}}{2002}]{Athana%
ssoula02}
{Athanassoula} E.,  2002, \apjl, 569, L83

\bibitem[\protect\citeauthoryear{{Athanassoula}}{{Athanassoula}}{2012}]{Athana%
ssoula2012}
{Athanassoula} E.,  2012, \mnras, 426, L46

\bibitem[\protect\citeauthoryear{{Baba}, {Saitoh} \& {Wada}}{{Baba}
  et~al.}{2013}]{Baba12}
{Baba} J.,  {Saitoh} T.~R.,    {Wada} K.,  2013, \apj, 763, 14

\bibitem[\protect\citeauthoryear{{Bertin} \& {Lin}}{{Bertin} \&
  {Lin}}{1996}]{Bertin96}
{Bertin} G.,  {Lin} C.~C.,  1996, Cambridge, MA MIT Press

\bibitem[\protect\citeauthoryear{{Binney} \& {Tremaine}}{{Binney} \&
  {Tremaine}}{2008}]{Binney08}
{Binney} J.,  {Tremaine} S.,  2008, Galactic Dynamics, 2nd edn. Princeton Univ.
  Press, Princeton, NJ

\bibitem[\protect\citeauthoryear{{Bird}, {Kazantzidis} \& {Weinberg}}{{Bird}
  et~al.}{2012}]{Bird12}
{Bird} J.~C.,  {Kazantzidis} S.,    {Weinberg} D.~H.,  2012, \mnras, 420, 913

\bibitem[\protect\citeauthoryear{{Buta}, {Byrd} \& {Freeman}}{{Buta}
  et~al.}{2004}]{Buta04}
{Buta} R.~J.,  {Byrd} G.,    {Freeman} T.,  2004, AJ, 127, 1982

\bibitem[\protect\citeauthoryear{{Buta}, {Laurikainen}, {Salo}, {Block} \&
  {Knapen}}{{Buta} et~al.}{2005}]{Buta05}
{Buta} R.~J.,  {Laurikainen} E.,  {Salo} H.,  {Block} D.~L.,    {Knapen}
  J.~H.~R.,  2005, AAS, 37, 1481

\bibitem[\protect\citeauthoryear{{Buta} \& {Shang}}{{Buta} \&
  {Shang}}{2011}]{Buta11}
{Buta} R.~J.,  {Shang} X.,  2011, MSAIS, 18, 13

\bibitem[\protect\citeauthoryear{{Byrd}, {Freeman} \& {Buta}}{{Byrd}
  et~al.}{2006}]{Byrd06}
{Byrd} R.,  {Freeman} T.,    {Buta} R.~J.,  2006, AJ, 131, 1377

\bibitem[\protect\citeauthoryear{{Clemens}, {Sanders} \& {Scoville}}{{Clemens}
  et~al.}{1988}]{Clemens88}
{Clemens} D.~P.,  {Sanders} D.~B.,    {Scoville} N.~Z.,  1988, \apj, 327, 139

\bibitem[\protect\citeauthoryear{{Col\'in}, {Avila-Reese},
  {V\'azquez-Semadeni}, {Valenzuela} \& {Ceverino}}{{Col\'in}
  et~al.}{2010}]{Colin11}
{Col\'in} P.,  {Avila-Reese} V.,  {V\'azquez-Semadeni} E.,  {Valenzuela} O.,
  {Ceverino} D.,  2010, \apj, 713, 535

\bibitem[\protect\citeauthoryear{{Col\'in}, {Valenzuela} \& {Klypin}}{{Col\'in}
  et~al.}{2006}]{Colinetal06}
{Col\'in} P.,  {Valenzuela} O.,    {Klypin} A.,  2006, \apj, 644, 687

\bibitem[\protect\citeauthoryear{{Comparetta} \& {Quillen}}{{Comparetta} \&
  {Quillen}}{2012}]{Comparetta12}
{Comparetta} J.,  {Quillen} A.~C.,  2012, ArXiv e-prints (arXiv:1207.5753)

\bibitem[\protect\citeauthoryear{{Di Matteo}, {Haywood}, {Combes}, {Semelin} \&
  {Snaith}}{{Di Matteo} et~al.}{2013}]{Dimatteo13}
{Di Matteo} P.,  {Haywood} M.,  {Combes} F.,  {Semelin} B.,    {Snaith} O.~N.,
  2013, ArXiv e-prints (arXiv:1301.2545)

\bibitem[\protect\citeauthoryear{{D'Onghia}, {Vogelsberger} \&
  {Hernquist}}{{D'Onghia} et~al.}{2013}]{Donghia12}
{D'Onghia} E.,  {Vogelsberger} M.,    {Hernquist} L.,  2013, \apj, 766, 14

\bibitem[\protect\citeauthoryear{{Ferreras}, {Cropper}, {Kawata} \& {Page}
  M.~{Erik}}{{Ferreras} et~al.}{2012}]{Ferreras12}
{Ferreras} I.,  {Cropper} M.,  {Kawata} D.,    {Page} M.~{Erik} A.,  2012,
  \mnras, 424, 1636

\bibitem[\protect\citeauthoryear{{Foyle}, {Rix}, {Doobs}, {Leroy} \&
  {Walter}}{{Foyle} et~al.}{2011}]{Foyle11}
{Foyle} K.,  {Rix} H.~W.,  {Doobs} C.~L.,  {Leroy} A.~K.,    {Walter} F.,
  2011, \apj, 735, 101

\bibitem[\protect\citeauthoryear{{Friedli} \& {Benz}}{{Friedli} \&
  {Benz}}{1993}]{Friedli93}
{Friedli} D.,  {Benz} W.,  1993, \aa, 268, 65

\bibitem[\protect\citeauthoryear{{Gerhard}}{{Gerhard}}{2011}]{Gerhard11}
{Gerhard} .,  2011, MSAIS, 18, 185

\bibitem[\protect\citeauthoryear{{Grand}, {Kawata} \& {Cropper}}{{Grand}
  et~al.}{2012a}]{Grand2012a}
{Grand} R.~J.~J.,  {Kawata} D.,    {Cropper} M.,  2012a, \mnras, 426, 167

\bibitem[\protect\citeauthoryear{{Grand}, {Kawata} \& {Cropper}}{{Grand}
  et~al.}{2012b}]{Grand2012b}
{Grand} R.~J.~J.,  {Kawata} D.,    {Cropper} M.,  2012b, \mnras, 421, 1529

\bibitem[\protect\citeauthoryear{{Hernquist}}{{Hernquist}}{1993}]{Hernquist93}
{Hernquist} L.,  1993, ApJS, 86, 389

\bibitem[\protect\citeauthoryear{{Hohl}}{{Hohl}}{1971}]{Hohl1971}
{Hohl} F.,  1971, \apj, 168, 343

\bibitem[\protect\citeauthoryear{{Klypin}, {Valenzuela}, {Col\'in} \&
  {Quinn}}{{Klypin} et~al.}{2009}]{Klypin2009}
{Klypin} A.~A.,  {Valenzuela} O.,  {Col\'in} P.,    {Quinn} T.,  2009, MNRAS,
  398, 1027

\bibitem[\protect\citeauthoryear{{Klypin}, {Zhao} \& {Somerville}}{{Klypin}
  et~al.}{2002}]{Klypin02}
{Klypin} A.~A.,  {Zhao} H.,    {Somerville} R.~S.,  2002, \apj, 573, 597

\bibitem[\protect\citeauthoryear{{Kravtsov}}{{Kravtsov}}{2003}]{Kravtsov2003}
{Kravtsov} A.~V.,  2003, \apj, 590, L1

\bibitem[\protect\citeauthoryear{{Kravtsov}, {Klypin} \& {Khokhlov}}{{Kravtsov}
  et~al.}{1997}]{Kravtsov1997}
{Kravtsov} A.~V.,  {Klypin} A.~A.,    {Khokhlov} A.~M.,  1997, ApJS, 111, 73

\bibitem[\protect\citeauthoryear{{Mart\'inez-Garc\'ia},
  {Gonz\'alez-L\'opezlira} \& {Bruzual}}{{Mart\'inez-Garc\'ia}
  et~al.}{2009}]{Martinez09}
{Mart\'inez-Garc\'ia} E.~E.,  {Gonz\'alez-L\'opezlira} R.~A.,    {Bruzual} G.,
  2009, \apj, 694, 512

\bibitem[\protect\citeauthoryear{{Martos}, {Hern\'andez}, {Y\'a\~nez}, {Moreno}
  \& {Pichardo}}{{Martos} et~al.}{2004}]{Martos04}
{Martos} M.,  {Hern\'andez} X.,  {Y\'a\~nez} M.,  {Moreno} E.,    {Pichardo}
  B.,  2004, \mnras, 350, 47

\bibitem[\protect\citeauthoryear{{Masset} \& {Tagger}}{{Masset} \&
  {Tagger}}{1997}]{Masset97}
{Masset} F.,  {Tagger} M.,  1997, \aa, 322, 442

\bibitem[\protect\citeauthoryear{{Meidt}, {Rand} \& {Merrifield}}{{Meidt}
  et~al.}{2009}]{Meidt09}
{Meidt} S.~E.,  {Rand} R.~J.,    {Merrifield} M.~R.,  2009, \apj, 702, 277

\bibitem[\protect\citeauthoryear{{Miller}, {Prendergast} \& {Quirk}}{{Miller}
  et~al.}{1970}]{Miller70}
{Miller} R.~H.,  {Prendergast} K.~H.,    {Quirk} W.~J.,  1970, \apj, 161, 903

\bibitem[\protect\citeauthoryear{{Miller} \& {Smith}}{{Miller} \&
  {Smith}}{1979}]{Miller1979}
{Miller} R.~H.,  {Smith} B.~F.,  1979, \apj, 227, 785

\bibitem[\protect\citeauthoryear{{Minchev}, {Famaey}, {Quillen}, {Di Matteo},
  {Combes}, {Vlaji\'c}, {Erwin} \& {Bland-Hawthorn}}{{Minchev}
  et~al.}{2012}]{Minchev12}
{Minchev} I.,  {Famaey} B.,  {Quillen} A.~C.,  {Di Matteo} P.,  {Combes} F.,
  {Vlaji\'c} M.,  {Erwin} P.,    {Bland-Hawthorn} J.,  2012, \aa, 548, 126

\bibitem[\protect\citeauthoryear{{Navarro}, {Frenk} \& {White}}{{Navarro}
  et~al.}{1997}]{Navarro97}
{Navarro} J.~F.,  {Frenk} C.~S.,    {White} S.~D.~M.,  1997, \apj, 490, 493

\bibitem[\protect\citeauthoryear{{Ostriker} \& {Peebles}}{{Ostriker} \&
  {Peebles}}{1973}]{Ostriker73}
{Ostriker} J.~P.,  {Peebles} P.~J.~E.,  1973, \apj, 186, 467

\bibitem[\protect\citeauthoryear{{Quillen}, {Dougherty}, {Bagley}, {Minchev} \&
  {Comparetta}}{{Quillen} et~al.}{2011}]{Quillen11}
{Quillen} A.~C.,  {Dougherty} J.,  {Bagley} M.~B.,  {Minchev} I.,
  {Comparetta} J.,  2011, \mnras, 417, 762

\bibitem[\protect\citeauthoryear{{Rautiainen} \& {Salo}}{{Rautiainen} \&
  {Salo}}{1999}]{Rautiainen99}
{Rautiainen} P.,  {Salo} H.,  1999, \aa, 348, 737

\bibitem[\protect\citeauthoryear{{Romero-G\'omez}, {Athanassoula}, {Antoja} \&
  {Figueras}}{{Romero-G\'omez} et~al.}{2011}]{Romerogomez11}
{Romero-G\'omez} M.,  {Athanassoula} E.,  {Antoja} T.,    {Figueras} F.,  2011,
  \mnras, 418, 1176

\bibitem[\protect\citeauthoryear{{Romero-G\'omez}, {Athanassoula}, {Masdemont}
  \& {Garc\'ia-G\'omez}}{{Romero-G\'omez} et~al.}{2007}]{Romero07}
{Romero-G\'omez} M.,  {Athanassoula} E.,  {Masdemont} J.~J.,
  {Garc\'ia-G\'omez} C.,  2007, \aa, 472, 63

\bibitem[\protect\citeauthoryear{{Roskar}, {Debattista}, {Loebman}, {Ivezi\'c}
  \& {Quinn}}{{Roskar} et~al.}{2011}]{Roskar11}
{Roskar} R.,  {Debattista} V.~P.,  {Loebman} S.~R.,  {Ivezi\'c} Z.,    {Quinn}
  T.~R.,  2011, ASPC, 448, 371

\bibitem[\protect\citeauthoryear{{Roskar}, {Debattista}, {Quinn}, {Stinson} \&
  {Wadsley}}{{Roskar} et~al.}{2008}]{Rok2008}
{Roskar} R.,  {Debattista} V.~P.,  {Quinn} T.~R.,  {Stinson} G.~S.,
  {Wadsley} J.,  2008, \apjl, 684, 79

\bibitem[\protect\citeauthoryear{{S{\'a}nchez}, {Kennicutt}, {Gil de Paz} \&
  {et al.}}{{S{\'a}nchez} et~al.}{2012}]{Sanchez2012}
{S{\'a}nchez} S.~F.,  {Kennicutt} R.~C.,  {Gil de Paz} A.,    {et al.} 2012,
  \aap, 538, A8

\bibitem[\protect\citeauthoryear{{Sanders} \& {Huntley}}{{Sanders} \&
  {Huntley}}{1976}]{Sanders76}
{Sanders} R.~H.,  {Huntley} J.~M.,  1976, \apj, 209, 53

\bibitem[\protect\citeauthoryear{{Sch\"onrich} \& {Binney}}{{Sch\"onrich} \&
  {Binney}}{2009}]{Schonrich09}
{Sch\"onrich} R.,  {Binney} J.,  2009, \mnras, 396, 203

\bibitem[\protect\citeauthoryear{{Sellwood}}{{Sellwood}}{2000}]{Sellwood00}
{Sellwood} J.~A.,  2000, Ap\&SS, 272, 31

\bibitem[\protect\citeauthoryear{{Sellwood} \& {Athanassoula}}{{Sellwood} \&
  {Athanassoula}}{1986}]{Sellwood86}
{Sellwood} J.~A.,  {Athanassoula} E.,  1986, \mnras, 221, 195

\bibitem[\protect\citeauthoryear{{Sellwood} \& {Binney}}{{Sellwood} \&
  {Binney}}{2002}]{Sellwood02}
{Sellwood} J.~A.,  {Binney} J.~J.,  2002, \mnras, 336, 785

\bibitem[\protect\citeauthoryear{{Sellwood} \& {Debattista}}{{Sellwood} \&
  {Debattista}}{2009}]{Sellwood09}
{Sellwood} J.~A.,  {Debattista} V.~P.,  2009, \mnras, 398, 1279

\bibitem[\protect\citeauthoryear{{Sellwood} \& {Sparke}}{{Sellwood} \&
  {Sparke}}{1988}]{Sparke1988}
{Sellwood} J.~A.,  {Sparke} L.~S.,  1988, \mnras, 231, 25

\bibitem[\protect\citeauthoryear{{Sheth} \& {Rossi}}{{Sheth} \&
  {Rossi}}{2010}]{Sheth10}
{Sheth} R.~K.,  {Rossi} G.,  2010, \mnras, 403, 2137

\bibitem[\protect\citeauthoryear{{Springel}}{{Springel}}{2005}]{Springel05}
{Springel} V.,  2005, \mnras, 364, 1105

\bibitem[\protect\citeauthoryear{{Toomre}}{{Toomre}}{1990}]{Toomre90}
{Toomre} A.,  1990, International Conference on Dynamics and Interactions of
  Galaxies, p.~292

\bibitem[\protect\citeauthoryear{{Tremaine} \& {Weinberg}}{{Tremaine} \&
  {Weinberg}}{1984}]{Tremaine84}
{Tremaine} S.,  {Weinberg} M.~D.,  1984, \apjl, 282, 5

\bibitem[\protect\citeauthoryear{{Tsoutsis}, {Kalapotharakos}, {Efthymiopoulos}
  \& {Contopoulos}}{{Tsoutsis} et~al.}{2009}]{Tsoutsis09}
{Tsoutsis} P.,  {Kalapotharakos} C.,  {Efthymiopoulos} C.,    {Contopoulos} G.,
   2009, \aa, 495, 743

\bibitem[\protect\citeauthoryear{{Valenzuela} \& {Klypin}}{{Valenzuela} \&
  {Klypin}}{2003}]{Valenzuela2003}
{Valenzuela} O.,  {Klypin} A.~A.,  2003, \mnras, 345, 406

\bibitem[\protect\citeauthoryear{{Widrow} \& {Dubinski}}{{Widrow} \&
  {Dubinski}}{2005}]{Widrow05}
{Widrow} L.~M.,  {Dubinski} J.,  2005, \apj, 631, 838

\bibitem[\protect\citeauthoryear{{Wielen}, {Fuchs} \& {Dettbarn}}{{Wielen}
  et~al.}{1996}]{Wielen1996}
{Wielen} R.,  {Fuchs} B.,    {Dettbarn} C.,  1996, \aa, 314, 438

\end{thebibliography}
\IfFileExists{\jobname.bbl}{}
{\typeout{}
\typeout{****************************************************}
\typeout{****************************************************}
\typeout{** Please run "bibtex \jobname" to optain}
\typeout{** the bibliography and then re-run LaTeX}
\typeout{** twice to fix the references!}
\typeout{****************************************************}
\typeout{****************************************************}
\typeout{}
}

\bsp

\label{lastpage}

\end{document}